\documentclass[12pt]{article}
\usepackage{instruction,epsfig}

\def\Journal#1#2#3#4{{#1} {\bf #2}, #3 (#4)}

\def\NIM{\em Nucl. Instrum. Methods}

\def\NPA{{\em Nucl. Phys.} A}
\def\NPB{{\em Nucl. Phys.} B}
\def\PLB{{\em Phys. Lett.}  B}
\def\PRL{\em Phys. Rev. Lett.}
\def\PRD{{\em Phys. Rev.} D}

\def \JHEP{{\em JHEP}}
\def \JETP{{\em JETP Lett.}}

\def\be{\begin{equation}}
\def\ee{\end{equation}}
\def\bea{\begin{eqnarray}}
\def\eea{\end{eqnarray}}

\def\kcnn{K^{\pm} \rightarrow \pi^\pm \pi^0 \pi^0}
\def\kccc{K^{\pm} \rightarrow \pi^\pm \pi^+ \pi^-}
\def\p0p0{\pi^0 \pi^0}
\def\pp{\pi^+ \pi^-}
\def\mmm{M_{00}}
\def\mm2{M_{00}^2}

\begin{document}
\title{\Large Observation of a cusp-like structure in the $\p0p0$ 
invariant mass distribution from $\kcnn$ decay and 
determination of the $\pi\pi$ scattering lengths}
\author{{\large The NA48/2 Collaboration}\  \\
J.R.~Batley,
C.~Lazzeroni,
D.J.~Munday,
M.W.~Slater,
S.A.~Wotton \\
{\em \small Cavendish Laboratory, University of Cambridge, Cambridge, CB3 0HE,
U.K.$\,$\footnotemark[1]} \\[0.2cm]
R.~Arcidiacono,
G.~Bocquet,
N.~Cabibbo,
A.~Ceccucci,
D.~Cundy$\,$\footnotemark[2],
V.~Falaleev,
M.~Fidecaro,
L.~Gatignon,
A.~Gonidec,
W.~Kubischta,
A.~Norton,
M.~Patel,
A.~Peters \\
{\em \small CERN, CH-1211 Geneva 23, Switzerland} \\[0.2cm]
S.~Balev,
P.L.~Frabetti,
E.~Goudzovski,
P.~Hristov$\,$\footnotemark[3],
V.~Kekelidze$\,$\footnotemark[3],
V.~Kozhuharov,
L.~Litov,
D.~Madigozhin,
E.~Marinova,
N.~Molokanova,
I.~Polenkevich,
Yu.~Potrebenikov,
S.~Stoynev,
A.~Zinchenko \\
{\em \small Joint Institute for Nuclear Research, Dubna, Russian    Federation} \\[0.2cm]
E.~Monnier$\,$\footnotemark[4],
E.~Swallow,
R.~Winston\\
{\em \small The Enrico Fermi Institute, The University of Chicago, Chicago, IL 60126, U.S.A.}\\[0.2cm]
 P.~Rubin,                      
 A.~Walker \\
{\em \small Department of Physics and Astronomy, University of
 Edinburgh, \\
JCMB King's Buildings, Mayfield Road, Edinburgh,    EH9 3JZ, U.K.} \\[0.2cm]
W.~Baldini,
A.~Cotta Ramusino,
P.~Dalpiaz,
C.~Damiani,
M.~Fiorini,  
A.~Gianoli,
M.~Martini,
F.~Petrucci,
M.~Savri\'e, 
M.~Scarpa,
H.~Wahl \\
{\em \small Dipartimento di Fisica dell'Universit\`a e Sezione    dell'INFN di Ferrara, I-44100 Ferrara, Italy} \\[0.2cm]
A.~Bizzeti$\,$\footnotemark[5],
M.~Calvetti,
E.~Celeghini, 
E.~Iacopini,
M.~Lenti,
F.~Martelli$\,$\footnotemark[6], \\
G.~Ruggiero$\,$\footnotemark[3],
M.~Veltri$\,$\footnotemark[6] \\
{\em \small Dipartimento di Fisica dell'Universit\`a e Sezione    dell'INFN di Firenze, I-50125 Firenze, Italy} \\[0.2cm]
M.~Behler,
K.~Eppard,
K.~Kleinknecht,
P.~Marouelli,
L.~Masetti,
U.~Moosbrugger,\\
C.~Morales Morales,
B.~Renk,
M.~Wache,
R.~Wanke,
A.~Winhart \\
{\em \small Institut f\"ur Physik, Universit\"at Mainz, D-55099 Mainz,
Germany$\,$\footnotemark[7]} \\[0.2cm]
D.~Coward$\,$\footnotemark[8],
A.~Dabrowski,
T.~Fonseca Martin$\,$\footnotemark[3],
M.~Shieh,
M.~Szleper,
M.~Velasco,
M.D.~Wood$\,$\footnotemark[9] \\
{\em \small Department of Physics and Astronomy,
Northwestern University, \\ Evanston, IL 60208-3112, U.S.A.}
 \\[0.2cm]
G.~Anzivino,
P.~Cenci,
E.~Imbergamo,
M.~Pepe,
M.C.~Petrucci,
M.~Piccini,
M.~Raggi,
M.~Valdata-Nappi \\
{\em \small Dipartimento di Fisica dell'Universit\`a e Sezione    dell'INFN di Perugia, I-06100 Perugia, Italy} \\[0.2cm]
C.~Cerri,
G.~Collazuol,
F.~Costantini,
L.~DiLella,
N.~Doble,
R.~Fantechi,
L.~Fiorini,
S.~Giudici,
G.~Lamanna,
I.~Mannelli,
A.~Michetti,
G.~Pierazzini,
M.~Sozzi \\
{\em \small Dipartimento di Fisica, Scuola Normale Superiore e Sezione
dell'INFN di Pisa, \\
I-56100 Pisa, Italy} \\[0.2cm]
B.~Bloch-Devaux,
C.~Cheshkov$\,$\footnotemark[3],
J.B.~Ch\`eze,
M.~De Beer,
J.~Derr\'e,
G.~Marel,
E.~Mazzucato,
B.~Peyaud,
B.~Vallage \\
{\em \small DSM/DAPNIA - CEA Saclay, F-91191 Gif-sur-Yvette, France} \\[0.2cm]
\newpage
M.~Holder,
A.~Maier$\,$\footnotemark[3],
M.~Ziolkowski \\
{\em \small Fachbereich Physik, Universit\"at Siegen, D-57068 Siegen,
Germany$\,$\footnotemark[10]} \\[0.2cm]
S.~Bifani,
C.~Biino,
N.~Cartiglia,
M.~Clemencic$\,$\footnotemark[3],
S.~Goy Lopez,
F.~Marchetto \\
{\em \small Dipartimento di Fisica Sperimentale dell'Universit\`a e
Sezione dell'INFN di Torino, \\
I-10125 Torino, Italy} \\[0.2cm]
H.~Dibon,
M.~Jeitler,
M.~Markytan,
I.~Mikulec,
G.~Neuhofer,
L.~Widhalm \\
{\em \small \"Osterreichische Akademie der Wissenschaften, Institut  f\"ur
Hochenergiephysik, \\
A-10560 Wien, Austria$\,$\footnotemark[11]} \\[1.5cm]
\rm
\setcounter{footnote}{0}
\footnotetext[1]{Funded by the U.K.    Particle Physics and Astronomy Research Council}
\footnotetext[2]{Present address: Istituto di Cosmogeofisica del CNR
di Torino, I-10133 Torino, Italy}
\footnotetext[3]{Present address: CERN, CH-1211 Geneva 23, Switzerland}
\footnotetext[4]{Also at Centre de Physique des Particules de Marseille, IN2P3-CNRS, Universit\'e
de la M\'editerran\'ee, Marseille, France}
\footnotetext[5] {Also Istituto di Fisica, Universit\`a di Modena, I-41100  Modena, Italy} 
\footnotetext[6]{Istituto di Fisica, Universit\`a di Urbino, I-61029  Urbino, Italy}
\footnotetext[7]{Funded by the German Federal Minister for Education
and research under contract 05HK1UM1/1}
\footnotetext[8]{Permanent address: SLAC, Stanford University, Menlo Park, CA 94025, U.S.A.}
\footnotetext[9]{Present address: UCLA, Los Angeles, CA 90024, U.S.A.}
\footnotetext[10]{Funded by the German Federal Minister for Research and Technology (BMBF) under contract 056SI74}
\footnotetext[11]{Funded by the Austrian Ministry for Traffic and Research under the contract GZ 616.360/2-IV GZ 616.363/2-VIII, and by the Fonds f\"ur Wissenschaft und Forschung FWF Nr.~P08929-PHY}

}
\begin{titlepage}
\maketitle
\abstract{We report the results from a study of a partial sample of
$\sim 2.3 \times 10^7$ $\kcnn$ decays recorded by the NA48/2 experiment at 
the CERN SPS, showing an anomaly in the $\p0p0$ invariant mass $(M_{00})$ 
distribution in the region around $M_{00}= 2m_+$, where $m_+$ is the charged 
pion mass. This anomaly, never observed in previous experiments, can be 
interpreted as an effect due mainly to the final state charge exchange 
scattering process $\pp \rightarrow \p0p0$ in $\kccc$ decay \cite{cabibbo1}.
It provides a
precise determination of $a_0 - a_2$, the difference between the $\pi\pi$
scattering lengths in the isospin $I=0$ and $I=2$ states. A best fit to a
rescattering model \cite{cabibbo2} corrected for isospin symmetry breaking 
gives $(a_0 - a_2)m_+ = 0.268 \pm 0.010 ~(stat.) ~\pm 0.004 ~(syst.)$, with
additional external uncertainties of $\pm 0.013$ from 
branching ratio and theoretical uncertainties. 
If the correlation between $a_0$ and $a_2$ predicted by chiral symmetry 
is taken into account, this result becomes 
$(a_0 - a_2)m_+ = 0.264 \pm 0.006 ~(stat.) \pm 0.004 ~(syst.) \pm
0.013 ~(ext.) $.}
\vspace{0.3 cm}
\begin{center}
\it{To be published in Physics Letters B}
\end{center}
\end{titlepage}         
\newpage
\section{Introduction}
The NA48/2 experiment at the CERN SPS is searching for direct CP violation in 
$K^\pm$ decay to three pions. The experiment uses simultaneous $K^+$ and $K^-$
beams with a momentum of $60$ GeV/c propagating along the same beam line. Data
have been collected in 2003-04, providing samples of $\sim 4 \times 10^9$ fully
reconstructed $\kccc$ and $\sim 10^8$ $\kcnn$ decays. Here we report
the results from a study of a partial sample of 
$\sim 2.3 \times 10^7$ $\kcnn$ decays recorded in 2003, showing an anomaly 
in the $\p0p0$ invariant mass $(M_{00})$ 
distribution in the region around $M_{00}= 2m_+$, where $m_+$ is the charged 
pion mass. This anomaly, never observed in previous experiments, can be 
interpreted as an effect due mainly to the final state charge exchange 
scattering process $\pp \rightarrow \p0p0$ in $\kccc$ decay \cite{cabibbo1}.
A best fit to a rescattering model \cite{cabibbo2} provides
a precise determination of  $a_0 - a_2$, the difference between the S-wave 
$\pi\pi$ scattering lengths in the isospin $I=0$ and $I=2$ states.

\section{Beam and detectors}
The two simultaneous beams are produced by $400$ GeV protons impinging on
a 40 cm long Be target. Particles of opposite charge with a central momentum of
$60$ GeV/c and a momentum band of $\pm 3.8\%$ produced at zero angle are
selected by a system of dipole magnets forming an ``achromat'' with null
total deflection, focusing quadrupoles, muon sweepers and collimators.
With $7\times 10^{11}$ protons per burst of $\sim 4.5$ s duration incident on 
the target the positive (negative) 
beam flux at the entrance of the decay volume
is  $3.8\times 10^{7}$ ($2.6\times 10^{7}$) particles per pulse, of which
$\sim 5.7\%$ ($\sim 4.9\%$) are $K^+$ ($K^-$). The decay volume is a 114 m 
long vacuum tank with a diameter of 1.92 m for the first 66 m, and 2.4 m 
for the rest.

Charged particles from $K^\pm$ decays are measured by a magnetic spectrometer 
consisting of four drift chambers \cite{augustin} and a large-aperture dipole
magnet located between the second and third chamber. Each chamber has eight 
planes of sense wires, two horizontal, two vertical and two along each of two
orthogonal $45^\circ$ directions. The spectrometer is located in a tank filled
with helium at atmospheric pressure  and separated from the decay volume by
a thin (0.0031 radiation lengths, $ X_0$) Kevlar window. A 16 cm diameter
vacuum tube centered on the beam axis runs the length of the spectrometer 
through central holes in the Kevlar window, drift chambers and calorimeters.
Charged particles are magnetically deflected in the horizontal plane by 
an angle corresponding to a transverse momentum kick of 
$120$ MeV/c. The momentum resolution of the spectrometer is 
$\sigma(p)/p = 1.02\% \oplus 0.044\%p$ ($p$ in GeV/c), as derived form the
known properties of the spectrometer and checked with the measured invariant
mass resolution of $\kccc$ decays. The magnetic spectrometer is followed by
a scintillator hodoscope consisting of two planes segmented into horizontal
and vertical strips and arranged in four quadrants.

A liquid Krypton calorimeter (LKr) \cite{Barr} is used to reconstruct 
$\pi^0 \rightarrow \gamma \gamma$ decays. It is an almost homogeneous 
ionization chamber with an active volume of 
$\sim 10 ~m^3$ of liquid krypton, segmented transversally into 13248 
$2~ cm \times 2 ~cm$ projective cells by a system of
Cu-Be ribbon electrodes, and with no longitudinal segmentation. 
The calorimeter is 27 $X_0$ thick and has an energy resolution 
$\sigma(E)/E = 0.032/\sqrt{E} \oplus 0.09/E \oplus 0.0042$ (E in GeV). 
The space resolution for single electromagnetic shower can be parametrized as
$\sigma_x = \sigma_y = 0.42/\sqrt{E} \oplus 0.06$ cm for each transverse 
coordinate $x,y$.

A neutral hodoscope consisting of a plane of scintillating fibers is installed
in the LKr calorimeter at a depth of $\sim 9.5 ~X_0$. It is
divided into four quadrants, each consisting of eight bundles of vertical
fibers optically connected to photomultiplier tubes.   

\section{Event selection and reconstruction}
The $\kcnn$ decays are selected by a two level trigger. 
The first level requires
a signal in at least one quadrant of the scintillator hodoscope in coincidence
with the presence of energy depositions in LKr consistent with at least two
photons. At the second level, a fast on-line processor receiving the drift
chamber information reconstructs the momentum of charged particles and
calculates the missing mass under the assumption that the particle is a 
$\pi^\pm$ originating from the decay of a 60 GeV/c $K^\pm$ travelling along 
the nominal beam axis.
The requirement that the missing mass is not consistent with the $\pi^0$ mass
rejects most of the main $K^\pm \rightarrow \pi^\pm \pi^0$ background. The
typical rate of this trigger is $\sim 15,000$ per burst.

Events with at least one charged particle track having a momentum above 
5 GeV/c, and at least four energy clusters in LKr, each consistent with 
a photon and above an energy threshold of 3 GeV, 
are selected for further analysis. In 
addition, the relative track and photon timings must be consistent with the 
same event within the experimental resolution ($\sim 1.5 ns$).
The distance between any two 
photons in LKr is required to be larger than 10 cm, 
and the distance between each photon and the impact point of any
track on LKr must exceed 15 cm. Fiducial cuts on the distance of each photon
from the LKr edges and centre are also applied in order to ensure 
full containment of the electromagnetic showers and to remove effects from the
beam pipe.
Finally, the distance between the charged particle track 
and the beam axis at the first drift chamber is required 
to be larger than 12 cm.

\begin{figure}
\begin{center}
\psfig{figure=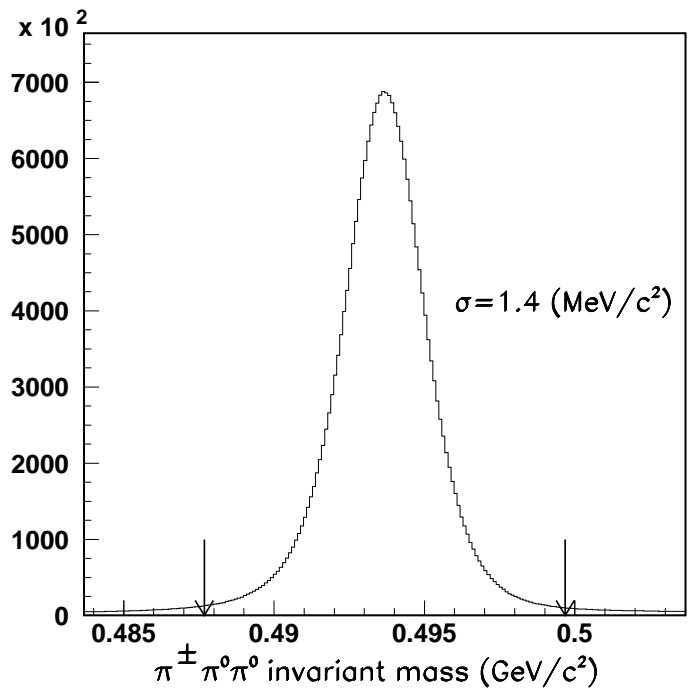,height=10.0cm}
\end{center}
\caption{ \it Invariant mass distribution of reconstructed $\pi^\pm \pi^0 \pi^0$ 
candidate events. The arrows indicate the selected mass interval.}
\label{fig:kmass}
\end{figure}

At the following step of the analysis we check the consistency of the surviving
events with the $\kcnn$ decay hypothesis. We assume that each possible pair of
photons originates from $\pi^0 \rightarrow \gamma \gamma$ decay and we calculate 
the distance $D_{ik}$ between the $\pi^0$ decay vertex 
and the LKr:
$$
D_{ik} = 
\frac
{\sqrt{E_i E_k [(x_i-x_k)^2+(y_i-y_k)^2 ]}}
{m_0}
$$   
where $E_i$,$E_k$ are the energies of the i-th and k-th photon, respectively,
$x_i,y_i,x_k,y_k$ are the coordinates of the impact point on LKr, and $m_0$ is
the $\pi^0$ mass. 
Among all photon pairs,
the two with the smallest $D_{ik}$
difference are selected as the best combination consistent with the two $\pi^0$
mesons from $\kcnn$ decay, and the distance of the $K^\pm$ 
decay vertex from the LKr is taken as the arithmetic average of the two 
$D_{ik}$ values (it can be
demonstrated that this choice gives the best $\p0p0$ invariant mass resolution
near threshold). Fig. \ref{fig:kmass} shows the invariant mass distribution 
of the system consisting of the two $\pi^0$ and a 
reconstructed charged particle track, assumed to be a $\pi^\pm$. 
This distribution is dominated by the $K^\pm$
peak, as expected. The non Gaussian tails originate from 
unidentified $\pi^\pm \rightarrow \mu^\pm$ in flight or wrong photon pairing.
The final event selection requires that the $\pi^\pm \pi^0 \pi^0$ 
invariant mass differs form the $K^\pm$ mass by at most $\pm 6$ MeV. 
This requirement is satisfied by $2.287\times 10^7$ events. 
The fraction of events with wrong photon pairing in this sample is 
$\sim 0.25 \%$, as estimated by a high-statistics fast
Monte Carlo simulation of $\kcnn$ decays which takes into account the momentum
distribution of the three pions, $\pi^0 \rightarrow \gamma \gamma$ decay 
kinematics and the effect of the detector acceptance and resolution.

\section{Cusp anomaly in the $\p0p0$ invariant mass distribution}
\begin{figure}
\begin{center}
\psfig{figure=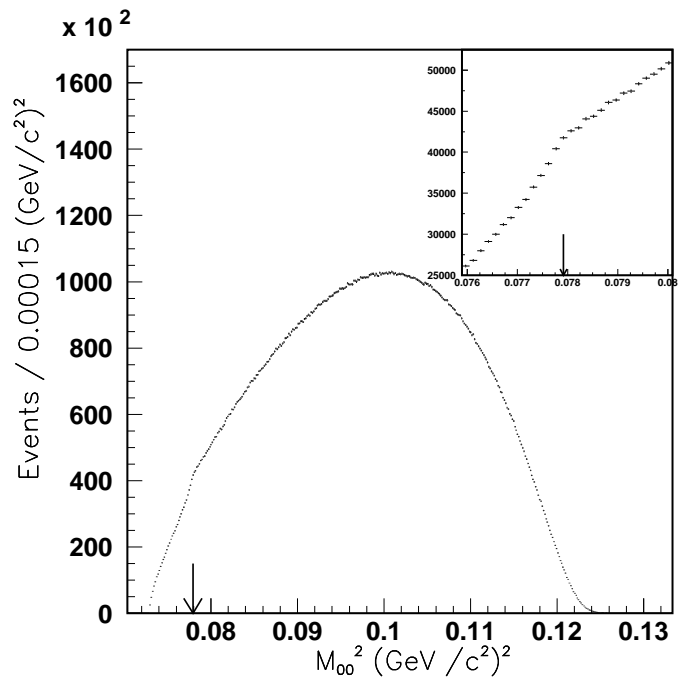,height=10.0cm}
\end{center}
\caption{ \it Distribution of $M_{00}^2$, the square of the $\p0p0$ 
invariant mass. The insert is an enlargement of a narrow region centered at
$\mm2 = (2m_+)^2$ (this point is indicated by the arrow). The statistical
error bars are also shown in these plots.}
\label{fig:moo}
\end{figure}

Figure \ref{fig:moo} shows the distribution of the square of the $\p0p0$ 
invariant mass, $\mm2$, for the final event sample. This distribution is 
displayed with a bin width of 0.00015 $(\mbox{GeV}/c^2)^2$, with the $51^{st}$
bin centered at $\mm2 = (2m_+)^2$ (as discussed below, the bin
width is chosen to be smaller than the $\mm2$ resolution).
A sudden change of slope near
$\mm2 = (2m_+)^2 = 0.07792~ (\mbox{GeV}/c^2)^2$ is clearly visible. 
Such an anomaly has not been observed in previous experiments.

The Dalitz plot distribution for $\kcnn$ decays is usually parametrized by a
series expansion in the Lorentz-invariant variable $u = (s_3-s_0)/m_+^2$,
where $s_i = (P_K - P_i)^2$ ($i$=1,2,3), 
$s_0 = (s_1+s_2+s_3)/3$, $P_K$ $(P_i)$
is the $K(\pi)$ four-momentum, and $i=3$ corresponds to the $\pi^\pm$ \cite{PDG}.
In our case $s_3 = \mm2$, and $s_0 = (m_K^2 + 2m_0^2+m_+^2)/3$. We have
used this parametrization in a fast Monte Carlo simulation of $\kcnn$ decays
with the same detector parameters used in previous NA48 analyses \cite{batley}.
This simulation takes into account most detector effects, including the trigger
efficiency and the presence of a small number $(<1\%)$ of ``dead'' LKr cells.
For any given value of the generated $\p0p0$ invariant mass the
simulation provides the detection probability and the distribution function for
the reconstructed value of $\mm2$. This allows the transformation of any
theoretical distribution into an expected distribution which can be compared 
directly with the measured one.

\begin{figure}
\begin{center}
\psfig{figure=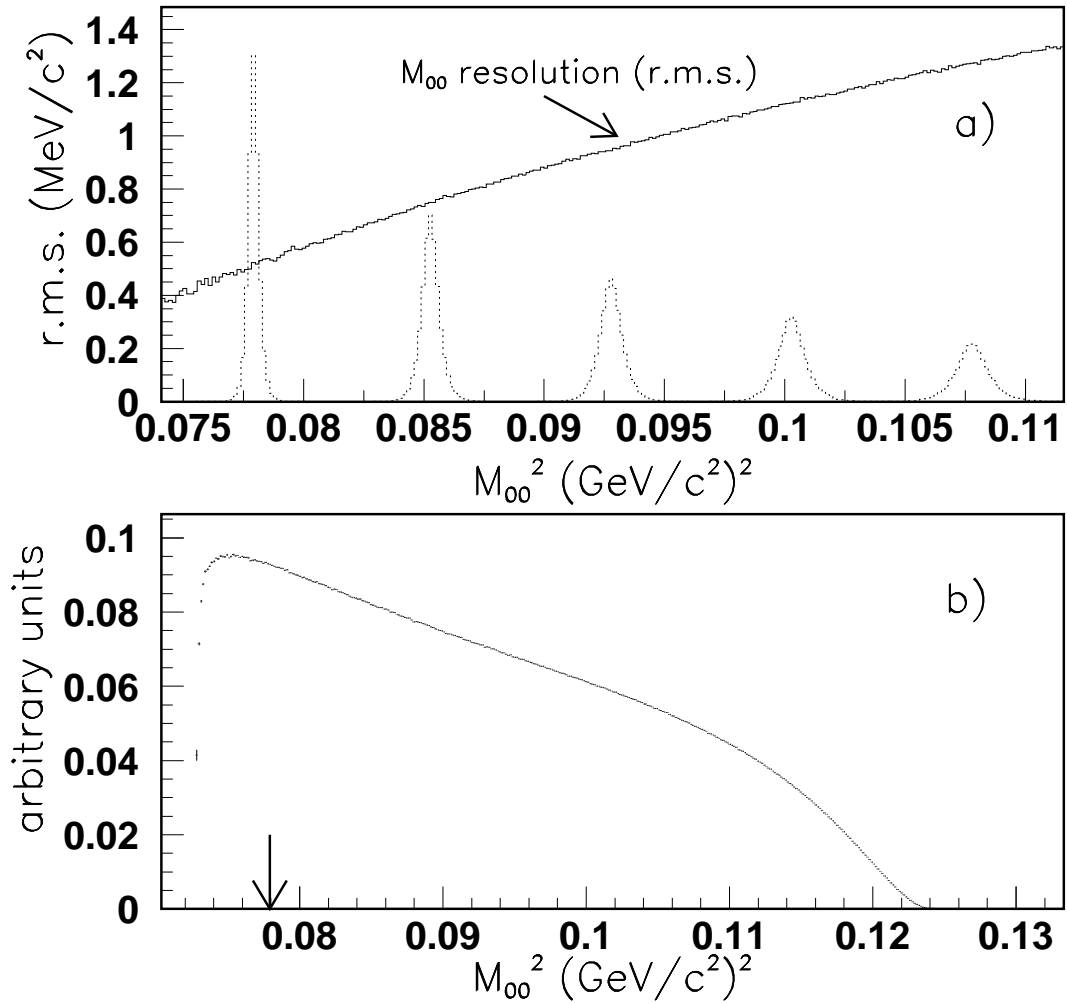,height=15.0cm}
\end{center}
\caption{ 
a) {\it Expected $M_{00}$ resolution (r.m.s. in $MeV/c^2$) versus generated
$\mm2$ (full line histogram), together with $\mm2$ distributions
for five generated values of $\mm2$};
b) {\it Acceptance versus $\mm2$ (see text). The point $\mm2 = (2m_+)^2$
is indicated by the arrow.}}
\label{fig:resolution}
\end{figure}

Figure \ref{fig:resolution}a shows the expected $\mmm$ resolution (r.m.s.)
as a function of $\mm2$, together with examples of $\mm2$ distributions
for five generated $\mm2$ values. The $\mmm$ resolution is the best
at small $\mm2$ values, varying between
$\sim 0.4  ~\mbox{MeV}/c^2$ near $\mmm = 2m_0$, and 
$\sim 1.4 ~\mbox{MeV}/c^2$ at the end of the $\mmm$ allowed range. 
It is $0.56 ~\mbox{MeV}/c^2$ at $\mmm = 2m_+$.
A plot of the overall detector
acceptance as a function of the generated $\mm2$ value, as predicted
by the Monte Carlo simulation (see Fig. \ref{fig:resolution}b),
shows no structure in the $\mm2$ region where the sudden change of
slope is observed in the data.

\begin{figure}
\begin{center}
\psfig{figure=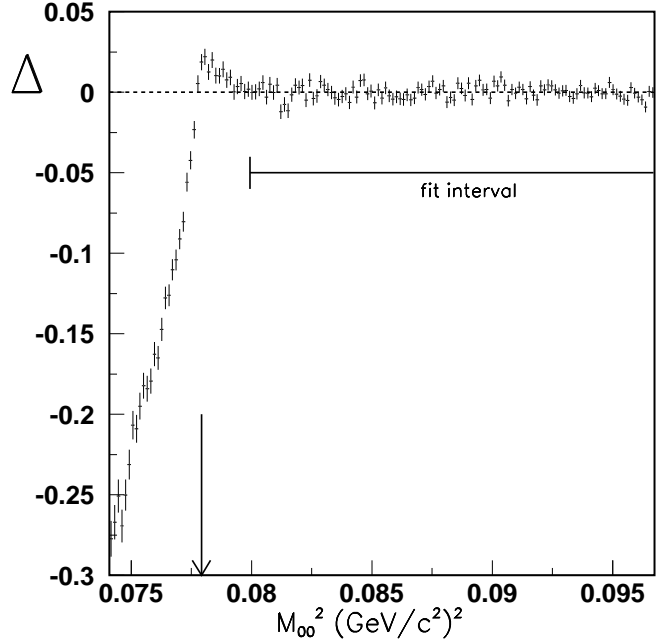,height=10.0cm}
\end{center}
\caption{ 
\it $\Delta \equiv $ (data - fit)/data versus $\mm2$. 
The point $\mm2 = (2m_+)^2$
is indicated by the arrow. Also shown is the $\mm2$ region used in the fit.
}
\label{fig:pdgfit}
\end{figure}

We have tried to fit the distribution of Fig. \ref{fig:moo} in the
interval $0.074<\mm2<0.097 ~ (\mbox{GeV}/c^2)^2$ using the distribution
predicted by the Monte Carlo simulation with a matrix element as given 
in ref. \cite{cabibbo1}:
\begin{equation}
{\cal M}_0 = 1 + \frac{1}{2}g_0u 
\label {eq:amp}
\end{equation}
In this fit the free parameters are $g_0$ and an overall normalization 
constant.
Because of the anomaly at $\mm2 = (2m_+)^2$, it is impossible to find a 
reasonable fit to the distribution of Fig. \ref{fig:moo} (the best fit gives
$\chi^2 = 9225$ for 149 degrees of freedom). However, fits with
acceptable $\chi^2$ values are obtained if the lower edge of the fit
interval is raised few bins above $\mm2 = (2m_+)^2$. As an example,
a fit in the interval $0.07994<\mm2<0.097 ~(\mbox{GeV}/c^2)^2$,
with the lower edge only 0.002 $(\mbox{GeV}/c^2)^2$ above $(2m_+)^2$,
gives $\chi^2 = 133.6$ for 110 degrees of freedom. This fit gives 
$g_0 = 0.683 \pm 0.001$ (statistical error only), in reasonable agreement with
the present world average, $g_0 = 0.638 \pm 0.020$ \cite{PDG} (it should be
noted, however, that the matrix element used here has not the same form as that
used in ref. \cite{PDG}). The quality of this fit is illustrated in Fig. 
\ref{fig:pdgfit}, which displays the quantity 
{\it $\Delta \equiv $ (data - fit)/data} 
as a function of $\mm2$ for the fit region 
$0.07994<\mm2<0.097 ~(\mbox{GeV}/c^2)^2$
and also for $\mm2 < 0.07994 ~ (\mbox{GeV}/c^2)^2$, where the prediction with 
the same parameters is extrapolated.

Figure \ref{fig:pdgfit} shows that, in the region $\mm2 < (2m_+)^2$, the data
fall below the prediction based on the same parameters obtained from the fit 
region. The total number of events in the first 50 bins of the data is 
$7.261 \times 10^5$, while the extrapolated prediction gives  
$8.359 \times 10^5$ events.

In order to investigate the origin of this ``deficit'' of events in the
data we have studied the event shape distributions 
in two 20 bins wide intervals, one just below and the other just above  
$\mm2 = (2m_+)^2$. 
Since $\mm2$ is computed using only information from the LKr calorimeter,
we consider only photon cluster parameters.
We denote the distributions of measured photon energy and distances 
in these two intervals as
$I_-$ and $I_+$, respectively, and we compare the $I_+/I_-$ ratios with those
predicted by the simulation after normalizing $I_-$ and $I_+$ to the 
same area.

\begin{figure}
\begin{center}
\psfig{figure=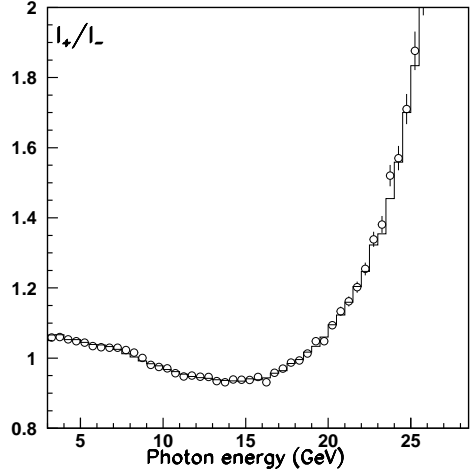,height=7.0cm}
\end{center}
\caption{ \it Data (points with error bars) - Monte Carlo (histogram) comparison
of the ratio of normalized photon energy distributions $I_+/I_-$ between 
events with $\mm2 > (2m_+)^2$ and $\mm2 < (2m_+)^2$ (see text)}
\label{fig:ratio1}
\end{figure}

\begin{figure}
\begin{center}
\psfig{figure=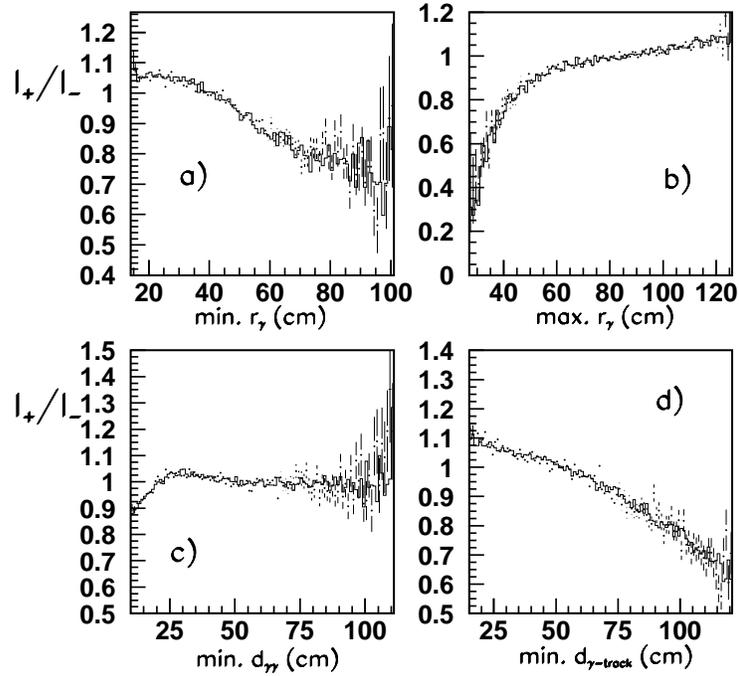,height=10.0cm}
\end{center}
\caption{ \it Data (points with error bars)  - Monte Carlo (histogram) 
comparison of ratios $I_+/I_-$ of normalized distance distributions 
between events with 
$\mm2 > (2m_+)^2$ and $\mm2 < (2m_+)^2$ (see text).
a) min. $r_\gamma$: distance (cm) between LKr centre and closest photon;
b) max. $r_\gamma$: distance (cm) between LKr centre and farthest photon; 
c) min. $d_{\gamma \gamma}$: minimum distance (cm) between photons at LKr;
d) $d_{\gamma-track}$: minimum distance (cm) between photons and tracks at LKr.
}
\label{fig:ratio2}
\end{figure}

These ratios (see Fig. \ref{fig:ratio1} and \ref{fig:ratio2}) show
that the shapes of all distributions for the two $\mm2$ intervals, as measured
in the data, are in excellent agreement with the Monte Carlo predictions. In
addition, no difference is observed between $K^+$ and $K^-$ nor between the 
data taken with opposite direction of the spectrometer magnetic field. 
The simulation also shows that the $\mm2$ distribution of the events affected 
by wrong photon pairing has no local structures over the whole $\mm2$ range. 
We conclude that the Monte Carlo simulation describes correctly 
the $\mm2$ dependence of the detection 
efficiency in the region around $\mm2 = (2m_+)^2$, and the ``deficit'' of
events in the data in the region  $\mm2 < (2m_+)^2$ is due to a real 
physical effect.

\section {Interpretation and determination of the $\pi\pi$ scattering lengths}

The sudden change of slope observed in the $\mm2$ distribution at 
$\mm2 = (2m_+)^2$ (see Fig. \ref{fig:moo}) suggests the presence of a 
threshold ``cusp'' effect from the decay $\kccc$ contributing to the $\kcnn$
amplitude through the charge exchange reaction 
$\pi^+ \pi^- \rightarrow \pi^0 \pi^0$. The presence of a cusp at 
$\mm2 = (2m_+)^2$ in $\p0p0$ elastic scattering due to the effect of virtual
$\pi^+ \pi^-$ loops has been discussed first by Meissner et al. \cite{meissner}.
For the case of $\kcnn$ decay Cabibbo has proposed a simple rescattering model
 \cite{cabibbo1}
describing the $\kcnn$ decay amplitude as the sum of two terms:
$$ 
{\cal M}(\kcnn) = {\cal M}_0 + {\cal M}_1  
$$
where ${\cal M}_0$ is the ``unperturbed amplitude'' of Eq. \ref{eq:amp},
and ${\cal M}_1$ is the contribution from the $\kccc$ decay amplitude through
$\pi^+ \pi^- \rightarrow \pi^0 \pi^0$
charge exchange, with the renormalization condition  ${\cal M}_1=0$ 
at $\mm2 = (2m_+)^2$. The contribution ${\cal M}_1$ is given by 
\begin{equation}
{\cal M}_1 = - 2 a_x m_+ {\cal M}_+\sqrt{1-\left(\frac{M_{00}}{2m_+}\right)^2}
\label {eq:amp1}
\end{equation}
where $a_x$ is the S-wave $\pi^+ \pi^-$ charge exchange scattering length
(threshold amplitude), and
${\cal M}_+$ is the known $\kccc$ decay amplitude at $M_{00}=2m_+$.
${\cal M}_1$ changes from real to imaginary at 
$M_{00} = 2m_+$ with the consequence that ${\cal M}_1$ interferes 
destructively with  ${\cal M}_0$ in the region $M_{00} < 2m_+$, while it
adds quadratically above it. In the limit of exact isospin symmetry
$a_x = (a_0-a_2)/3$, where $a_0$ and $a_2$ are the S-wave $\pi\pi$ 
scattering lengths in the $I=0$ and $I=2$ states, respectively.

\begin{figure}
\hspace*{-2.0 cm}   
\psfig{figure=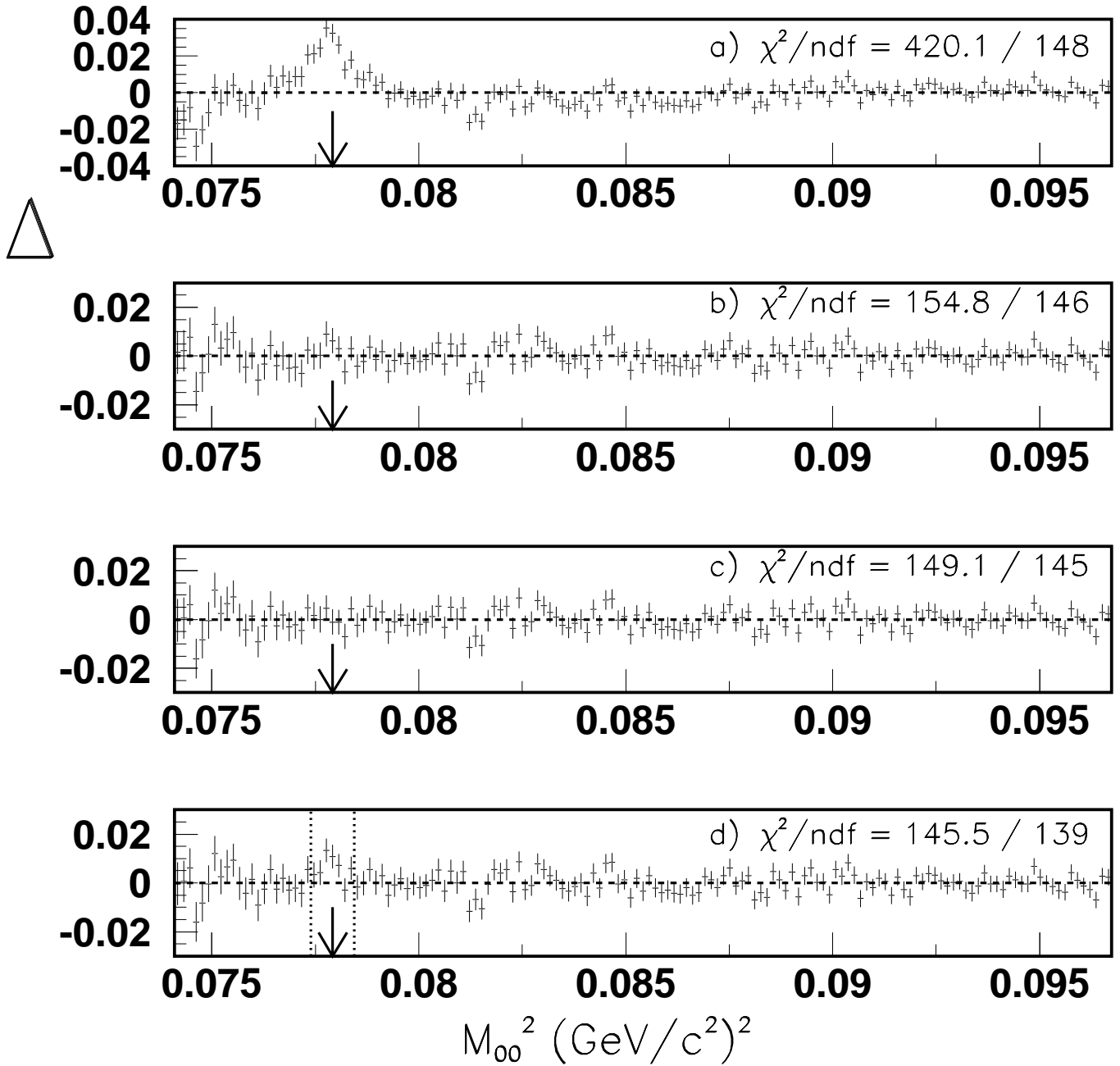,height=19.0cm}
\caption 
{\it $\Delta =$ (data - fit)/data versus $\mm2$ for various theoretical models:
a) using the simple charge-exchange model of ref. \cite{cabibbo1};  
b) fit to the rescattering model of ref. \cite{cabibbo2};
c) fit to the model of ref. \cite{cabibbo2} including pionium formation;
d) fit to the model of ref. \cite{cabibbo2} excluding a 7 bin wide interval 
centred at $\mm2 = (2m_+)^2$. The two vertical dotted lines in d) show the interval
excluded form the fit. The point $\mm2 = (2m_+)^2$  is indicated by the arrow.}

\label{fig:variousfit}
\end{figure}

In this simple rescattering model there is only one additional
parameter, $a_x m_+$. A fit to the $\mm2$ distribution in the interval
$0.074<\mm2<0.097 ~ (\mbox{GeV}/c^2)^2$ using $a_x m_+$ as a free parameter
gives $\chi^2 = 420.1$ for 148 degrees of freedom. The quality of this fit is
illustrated in Fig. \ref{fig:variousfit}a which displays the quantity 
$\Delta$ defined in section 4 as a function of $\mm2$. 
One can see that this model provides a much better but still 
unsatisfactory description of the data. 
In particular, the data points are systematically above the fit in the region 
near  $\mm2 = (2m_+)^2$.

Recently, Cabibbo and Isidori \cite{cabibbo2} have proposed a more complete 
formulation of the model which takes into account all rescattering processes
at the one-loop and two-loop level. In this formulation the matrix element for
$\kcnn$ decay includes several additional terms which 
depend on five S-wave scattering lengths, denoted by 
$a_x$, $a_{++}$, $a_{+-}$, $a_{+0}$, $a_{00}$, and describing 
$
\pi^+ \pi^- \rightarrow \pi^0 \pi^0,
 \pi^+ \pi^+ \rightarrow \pi^+ \pi^+,
 \pi^+ \pi^- \rightarrow \pi^+ \pi^-,
 \pi^+ \pi^0 \rightarrow \pi^+ \pi^0,
 \pi^0 \pi^0 \rightarrow \pi^0 \pi^0
$ 
scattering, respectively. In the limit of exact 
isospin symmetry these scattering lengths can all be expressed 
as linear combinations of $a_0$ and $a_{2}$. 

At tree level, omitting one-photon exchange diagrams, isospin symmetry 
breaking contributions to the elastic $\pi\pi$ scattering amplitude
can be expressed as a function of one parameter 
$ \epsilon = (m_+^2-m_0^2)/m_+^2 = 0.065$ \cite{kolck}.
In particular, the ratio between the threshold amplitudes
$a_x$, $a_{++}$, $a_{+-}$, $a_{+0}$, $a_{00}$ and the corresponding
isospin symmetric ones - evaluated at the $\pi^\pm$ mass - is
equal to
$1-\epsilon $ for 
$
\pi^+ \pi^+ \rightarrow \pi^+ \pi^+,
 \pi^+ \pi^0 \rightarrow \pi^+ \pi^0,
 \pi^0 \pi^0 \rightarrow \pi^0 \pi^0
$; 
$1+\epsilon $ for 
$\pi^+ \pi^- \rightarrow \pi^+ \pi^-$;
and 
 $1+\epsilon/3 $ for 
$\pi^+ \pi^- \rightarrow \pi^0 \pi^0$.
These corrections have been applied to the rescattering model of ref. 
\cite{cabibbo2} 
in order to extract $a_0$ and $a_2$ from the fit to the data.
 
In the model of ref. \cite{cabibbo2} the matrix element for $\kcnn$ 
decay includes terms which depend on both independent kinematic variables 
($M_{00}$ and  $M_{+0}$, the invariant mass of the $\pi^{\pm} \pi^0$ pair)  
requiring, therefore, a fit to the two-dimensional Dalitz plot. 
We have performed an approximate fit to this model by calculating these terms
at the average value of $M_{+0}^2$ for each value of $\mm2$.
This fit has five free parameters: 
$(a_0 - a_2)m_+, a_2m_+, g_0$, a quadratic term of the form $0.5 h^\prime u^2$ 
added in equation \ref{eq:amp} and an overall normalization constant.
The quality of the fit  ($\chi^2 = 154.8$ for 146 degrees of freedom)
is shown in Fig. \ref{fig:variousfit}b. 
A better fit ($\chi^2 = 149.1$ for 145 degrees of freedom, 
see Fig. \ref{fig:variousfit}c) is obtained by adding to the model 
a term describing the expected formation of $\pi^+\pi^-$ atoms (``pionium'') 
decaying to $\p0p0$ at $M_{00} = 2m_+$. 
The best fit value for the rate of $K^{\pm} \rightarrow \pi^\pm + $pionium 
decay, normalized to the $\kccc$ decay rate, 
is $(1.61 \pm 0.66)\times 10^{-5}$, in reasonable agreement
with the predicted value $\sim 0.8\times 10^{-5}$ \cite{silagadze}.  

The rescattering model of ref. \cite{cabibbo2} does not include radiative 
corrections, which are particularly important near $M_{00} = 2m_+$, 
and contribute to the formation of  $\pi^+\pi^-$ atoms. 
For this reason we prefer to exclude from the final fit a group of seven 
consecutive bins centered at $M_{00} = 2m_+$. The quality of this fit 
($\chi^2 = 145.5$ for 139 degrees of freedom) is illustrated in Fig. 
\ref{fig:variousfit}d,
which shows the small excess of events from pionium formation in the bins 
excluded from the fit.
Table \ref{tab:param} lists the best fit values of the parameters, as obtained by two
independent analyses which use different event selection criteria and different Monte Carlo
simulations to take into account acceptance and resolution effects (the analysis described so far
is denoted as Analysis A; analysis B uses a simulation of the detector
based on GEANT \cite{geant}).
We take the arithmetic average of these values as the measurement of these parameters, and one 
half of the difference between the two values as a systematic uncertainty from acceptance 
calculations. In both analyses changing the selection criteria never leads to variations of
the best fit parameters larger than these uncertainties. 

\begin{table}[here]
\begin{center}
\begin{tabular}{c c c c}
Parameter & Analysis A & Analysis B & Arithmetic average \\
\hline
$(a_0-a_2)m_+$   & $0.269 \pm 0.010$  & $0.268 \pm 0.010$  & $0.268 \pm 0.010$ \\
\hline
$a_{2}m_+$ & $-0.053 \pm 0.020$ & $-0.030 \pm 0.022$ & $-0.041 \pm 0.022$ \\
\hline
$g_0$        & $0.643 \pm 0.004$  & $0.647 \pm 0.004$ & $0.645 \pm 0.004$ \\
\hline
$h^\prime$   & $-0.055 \pm 0.010$  & $-0.039 \pm 0.012$ & $-0.047 \pm 0.012$ \\
\hline

\end{tabular}
\end{center}
\caption{\it Parameter best fit from two 
independent analyses (statistical error only)}
\label{tab:param}
\end{table}

\section{Other systematic uncertainties on the best fit parameters}
In addition to the systematic uncertainties associated with differences of the two
analyses, the following potential sources of systematic errors have been considered 
(see Table: \ref{tab:syst})

\subsection{Variation of the trigger efficiency over the $\mm2$ fit interval}
The trigger efficiency has been measured using a sample of ``minimum bias'' events 
recorded continuously by a trigger requiring only the presence of a signal in at least
two quadrants of the neutral hodoscope (during data taking the rate of this trigger was
downscaled by a large factor). Within statistical errors the dependence of the trigger 
efficiency on $\mm2$ is found to be consistent with the constant value 
$\epsilon_{tr} = 0.928 \pm 0.001$ for $(2m_0)^2 < \mm2 < 0.097 ~(\mbox{GeV}/c^2)^2$.
An equally good fit to the trigger efficiency is obtained using a 
$3^{rd}$ degree polynomial. Varying the polynomial
coefficients, so that the $\chi^2$ 
increases by an amount corresponding to 
$\pm 1\sigma$, $(a_0-a_2)m_+,a_2 m_+,g_0$ and $h^\prime$ change
as shown in Table \ref{tab:syst}.

\subsection{Dependence on the upper edge of the fit interval}
 The upper edge of the $\mm2$ fit interval has been varied from 0.094 to 0.107 
$~(\mbox{GeV}/c^2)^2$, resulting in variations of the best fit parameters with respect to the
default upper bound $\mm2 = 0.097 ~(\mbox{GeV}/c^2)^2$ as listed in table   
\ref{tab:syst}.

\subsection{Dependence on the position of the $K^\pm$ decay vertex}
As an additional check of the acceptance calculation,
the $\kcnn$ events have been subdivided into two independent samples with the distance $D$
of the reconstructed $K^\pm$ decay vertex from the LKr in the intervals
$48<D<88$ m, and $88<D<136$ m, respectively. The best fit parameter values obtained from
separate fits to the the two samples agree within statistics, providing no evidence for a
possible systematic uncertainty associated with the position of the kaon decay vertex.

\subsection{Dependence on the $K^\pm$ charge sign}
The $\kcnn$ events consist of $1.470\times 10^7$ $K^+$ and $0.817\times 10^7$ $K^-$.
Separate fits to these two samples give statistically consistent values for all best fit
parameters. The two values of the slope parameter $g_0$ are $g_0 = 0.638 \pm 0.005$ 
for $K^+$ and  $g_0 = 0.653 \pm 0.006$ for $K^-$, which disagree by $\sim 1.9\sigma$. 
We take one half of their difference (0.008) as a systematic uncertainty on the value of $g_0$ obtained by 
the fit to the full $K^\pm$ sample.

\subsection{Dependence on the distance between the $\pi^\pm$ track and the nearest
photon}
The $\pi^\pm$ interaction in LKr may produce multiple energy clusters which are 
located, in general, near the impact point of the $\pi^\pm$ track and in some cases
may be identified as photons. In order to study the effect of these fake photons on the best 
fit parameters we have repeated the analysis by varying the cut on the minimum distance $d$
between each photon and the track impact point on LKr (both analyses A and B require $d>15$ cm).
Varying $d$ between 10 and 25 cm changes $(a_0-a_2)m_+$ by $\pm 0.002$, while leaving the other
parameters unchanged. We take this variation as a systematic uncertainty on $(a_0-a_2)m_+$. 

\subsection{Effect of LKr resolution and non-linear response at low photon energies} 
The effect of possible uncertainties in the parameters describing the LKr energy resolution
has been simulated by adding a Gaussian noise with r.m.s. value of 0.06 GeV to the measured
photon energies. An additional uncertainty may arise form the correction applied to the measured
photon energies to account for the LKr non-linear response at low photon energies (typically
$<2\%$ at 3 GeV and becoming negligible above 10 GeV). The parameters describing this correction
have been varied within limits chosen so that the measured $\pi^0$ mass for symmetric photon
pairs does not depend on the $\pi^0$ energy. Varying both the LKr
resolution and non-linearity
correction parameters according to these procedures changes $(a_0-a_2) m_+$ by $\pm 0.001$, while
leaving the other parameters unchanged. We take this variation as an additional systematic
uncertainty on $(a_0-a_2)m_+$

Table \ref{tab:syst} lists all the systematics uncertainties discussed above. These are
added in quadrature to obtain the total experimental systematic error on the values of the
best fit parameters.
\begin{table}[here]
\begin{center}
\begin{tabular}{c| c| c| c| c| c| c| c}
 
  & {\small Acceptance} & {\small Trigger}  & {\small Fit} 
  & {\small $K^+/K^-$} & {\small $\pi^\pm - \gamma$} & {\small LKr} & {\small Total}\\
  & {\small calculation} & {\small efficiency}  & {\small interval} 
  & {\small difference} & {\small distance} & {\small response} & \\
  {\small Parameter} &             &             &            &            &          & & \\
  \hline
  $(a_0-a_2)m_+  $&$ 0.001$&$ 0.001 $&$ 0.0025 $&$ -    $&$ 0.002 $&$ 0.001 $&$ \pm 0.004$  \\
  \hline
  $a_2m_+        $&$ 0.012$&$ 0.005 $&$ 0.006 $&$ -     $&$ -      $&$ -    $&$ \pm 0.014$  \\
  \hline
  $g_0           $&$ 0.002$&$ 0.002 $&$ 0.002 $&$ 0.008 $&$ -      $&$ -    $&$ \pm 0.009$ \\
  \hline
  $h^\prime      $&$ 0.009$&$ 0.003 $&$ 0.006 $&$ -     $&$ -      $&$ -    $&$ \pm 0.011$ \\
  \hline
\end{tabular}
\end{center}
\caption{\it Systematic uncertainties}
\label{tab:syst}
\end{table}

\subsection{``External'' uncertainties}
A crucial parameter in the model of refs. \cite{cabibbo1} \cite{cabibbo2} is the ratio
$ R = A_{++-}/A_{+00}$ between the weak amplitudes of $\kccc$ and $\kcnn$ decay. The value extracted
from the measured decay branching ratio \cite{PDG} is $ R = 1.972 \pm 0.023$. Varying R withing its error
changes $(a_0-a_2)m_+$ by $\pm 0.003$, while leaving the other parameters unchanged. An additional
theoretical error of $\pm5\%$ on  $(a_0-a_2)m_+$, or $\pm 0.013$ is estimated in ref. \cite{cabibbo2} 
as the result of neglecting higher-order terms and radiative corrections in the rescattering model. 
These uncertainties have no significant effect on $a_2m_+$.

Taking into account all systematic and external uncertainties we quote:
\begin{eqnarray}
(a_0-a_2) m_+ = 0.268 \pm 0.010 \mbox{(stat)} \pm 0.004 \mbox{(syst)} \pm 0.013 \mbox{(ext)} 
\label{eq:res1} \\
a_2 m_+ = -0.041 \pm 0.022 \mbox{(stat)} \pm 0.014 \mbox{(syst)} \label{eq:res2} 
\end{eqnarray}
The two statistical errors from the fit are strongly correlated, with a correlation coefficient
of -0.858. We note that this analysis offers the first direct determination of $a_2$, though not
as precise as that of $a_0-a_2$.

Preliminary results obtained under the assumption of exact isospin symmetry have been reported earlier
\cite{giudici}.

\section{Fit using the correlation between $a_0$ and $a_2$ predicted by chiral symmetry}
 It has been shown that analyticity and chiral symmetry provide a constraint between
$a_0$ and $a_2$ \cite {colangelo1}:
$$
a_2 m_+ = (-0.0444\pm 0.0008) + 0.236(a_0m_+ -0.22) -0.61(a_0m_+ -0.22)^2 -9.9(a_0m_+ -0.22)^3
$$
Using this constraint in the fit to the rescattering model of ref. \cite{cabibbo2} we obtain
\begin{equation}
a_0m_+ = 0.220 \pm 0.006 \mbox{(stat)} \pm 0.004 \mbox{(syst)}\pm 0.011 \mbox{(ext)} \label{eq:res3} \\
\end{equation}
which corresponds to 
\begin{equation}
(a_0-a_2) m_+ = 0.264 \pm 0.006\mbox{(stat)} \pm 0.004 \mbox{(syst)} \pm 0.013 \mbox{(ext.)} 
\label{eq:res4} \\
\end{equation}

\section{Summary and conclusions}
The $\pi^0\pi^0$ invariant mass ($M_{00}$) distribution measured from a sample of 
$2.287 \times 10^7$ $\kcnn$ fully reconstructed decays collected by the NA48/2 experiment at the CERN
SPS shows an anomaly at $M_{00} = 2m_+$. This anomaly has been observed for the first time in this
experiment thanks to the large statistical sample and the excellent $M_{00}$ resolution. It can be
described by a rescattering model \cite{cabibbo1}  \cite{cabibbo2} dominated by the contribution 
from the decay $\kccc$ through the charge-exchange reaction $\pi^+ \pi^- \rightarrow \pi^0 \pi^0$.
These data have been used, therefore, to determine the difference $a_0-a_2$ between the $I=0$ and
$I=2$  S-wave $\pi\pi$ scattering lengths. Our result (see eq. \ref{eq:res1}) is in very good
agreement with theoretical calculations performed in the framework of Chiral Perturbation Theory
(ChPT) \cite {colangelo2}, which predict $(a_0-a_2)m_+ = 0.265 \pm 0.004$.
A different theoretical calculation based on a direct analysis of
$\pi\pi$
scattering data without using chiral symmetry \cite{pelaez} leads to a
somewhat different value with a larger
uncertainty, $(a_0-a_2)m_+ = 0.278 \pm 0.016$, which also agrees with
our result.

Previous determination of the $\pi\pi$ scattering lengths have relied on a variety of methods, such
as the measurement of $K^+ \rightarrow \pi^+ \pi^- e^+ \nu_e $ decay \cite{pislak}, also being 
studied by the NA48/2 collaboration, or the measurement of the lifetime of the $\pi^+\pi^-$ atom 
\cite{adeva}. Our value of $a_0$ (see eq. \ref{eq:res3}) is in good agreement with the result of
experiment 865 at BNL \cite{pislak}, 
$a_0m_+ = 0.216 \pm 0.013 \mbox{(stat)} \pm 0.002 \mbox{(syst)}\pm 0.002 \mbox{(theor.)}$,
also obtained using constraints based on analyticity and chiral symmetry. 
Our value of $a_0-a_2$ is also in good agreement with the first measurement of the lifetime
of the $\pi^+ \pi^-$ atom \cite{adeva}, which corresponds to $|a_0-a_2|m_+ = 0.264^{+0.033}_{-0.020}$
(it should be noted that the latter result provides only a determination of $|a_0-a_2|$, while our
measurement of $\kcnn$ decays is also sensitive to its sign).

To conclude, the study of a large sample of $\kcnn$ decays with excellent resolution on the
$\pi^0\pi^0$ invariant mass has provided a novel, precise determination of $a_0-a_2$, independent
of other methods and with different systematics uncertainties. In the near future the expected increase 
of the event sample by about a factor of 5 from the analysis of all the 2003-2004 data will further
reduce the statistical error of our measurement. To be useful, this will require an
improvement of the rescattering model to include higher-order terms and also radiative corrections.

\section*{Acknowledgments}        
We gratefully acknowledge the CERN SPS accelerator and beam-line staff for the excellent performance
of the beam. We thank the technical staff of the participating laboratories and universities for
their effort in the maintenance and operation of the detectors, and in data processing. It is also 
a pleasure to thank G. Colangelo, J. Gasser, B. Kubis and
A. Rusetsky for illuminating discussions and help on the subject of isospin symmetry breaking 
corrections. We also thank G. Isidori for valuable discussions on the fitting procedure.

\end{document}